\documentclass[12pt]{article}
\usepackage[utf8]{inputenc}

\usepackage{amsmath}
\usepackage{amsfonts}
\usepackage{amssymb}
\usepackage{makeidx}
\usepackage{graphicx}
\usepackage{subcaption}
\usepackage{listings}
\usepackage{apacite}
\usepackage{natbib}
\usepackage{indentfirst}
\usepackage{caption}

\usepackage{ulem}
\usepackage{soul}
\usepackage{color}

\usepackage{tabularx}
\usepackage{siunitx}
\usepackage{multirow}
\usepackage{booktabs}
\newcolumntype{Y}{>{\centering\arraybackslash}X}

\usepackage{setspace}
\usepackage{authblk}
\usepackage[pdfborder={0 0 0}]{hyperref}

\title{\bf Option Pricing with Mixed L\'{e}vy Subordinated Price Process and  Implied Probability Weighting Function  }
\author[a]{Abootaleb Shirvani}
\author[b]{Yuan Hu}
\author[c]{Svetlozar T. Rachev}
\author[d]{Frank J. Fabozzi}

\affil[a]{\small Department of Mathematics and Statistics, Texas Tech University\\
	\url{abootaleb.shirvani@ttu.edu}}
\affil[b]{\small Department of Mathematics and Statistics, Texas Tech University\\
	\url{yuan.hu@ttu.edu}}
\affil[c]{\small Department of Mathematics and Statistics, Texas Tech University\\
	\url{zari.rachev@ttu.edu}}
\affil[d]{\small EDHEC Business School\\
	\url{frank.fabozzi@edhec.edu}}

\date{}
\begin{document}

\begin{spacing}{1.00}
\thispagestyle{plain}
\maketitle

\bigbreak
\bigbreak

\noindent \textbf{Abstract}\ \ \ \ \ 
It is essential to incorporate the impact of investor behavior when modeling the dynamics of asset returns. In this paper, we reconcile behavioral finance and rational finance by incorporating investor behavior within the framework of dynamic asset pricing theory. To include the views of investors, we employ the method of subordination which has been proposed in the literature by including business (intrinsic, market) time.  We define a mixed L\'{e}vy subordinated model by adding a single subordinated L\'{e}vy process to the well-known log-normal model, resulting in a new log-price process. We apply the proposed models to study the behavioral finance notion of ``greed and fear" disposition from the perspective  of rational dynamic asset pricing theory. The greedy or fearful disposition of option traders is studied using the shape of the probability weighting function. We then derive the implied probability weighting function for the fear and greed deposition of option traders in comparison to spot traders. Our result shows the diminishing sensitivity of option traders. Diminishing sensitivity results in option traders overweighting the probability of big losses in comparison to spot traders. 

\newpage
\doublespacing
\section*{}
Several  studies provide empirical evidence that the behavior of investors has an impact on stock returns.\footnote{See, for example, \cite{Brown:2004},  \cite{Baker:2007}, and \cite{Long:1990}.} 
To obtain a more realistic log return pricing model, it is essential to incorporate investor behavior and investor sentiment.  \cite{Shefrin:2005} combined two different normally distributed log returns to represent the views of the buyer and seller for pricing options of a certain asset return model. The asset return model that he used, the mixture of normal distributions, is not infinity divisible due to its underlying finite support. Thus, according to \cite{Black:1973} and \cite{Merton1973a}, this model would lead to arbitrage opportunities, making it inappropriate for pricing options.

In rational finance, some researchers have modeled the price process by incorporating a subordinator process into the classical Black-Scholes-Merton (BSM) model. The subordinating, time change, process is a technique for introducing additional parameters into the return model for the purpose of capturing the following features: (1) the asymmetry and leptokurtic behavior of asset return distributions, (2) the effect of investor behavior and investor sentiment on the market underlying price model, (3) time-varying volatility of asset returns, (4) regime switching in stock market returns, and (5) leverage effects. 

\cite{Mandelbrot:1967} and \cite{Clark:1973} applied the concept of time change to the Brownian motion process to obtain a more realistic speculative price process. \cite{Merton:1976} introduced a jump-diffusion model using a compound Poisson time-change L\'{e}vy process. Two decades later,  \cite{Hurst:1997} applied various subordinated log return model processes to model the well-documented heavy-tail phenomena exhibited by asset return distributions.  The views of investors can be incorporated into log return asset pricing models and option pricing models by introducing an intrinsic time process, which is referred to as a behavioral subordinator \citep[see][]{Shirvani:2019}.

In this paper, we attempt to reconcile behavioral finance and rational finance by incorporating investor behavior into the framework of a dynamic asset pricing model. We extend the approach of \cite{Black:1973} and \cite{Merton1973a} by mixing a subordinated L\'{e}vy process, with a Gaussian component to represent investor behavior. The price process -- referred to as a mixed L\'{e}vy subordinated market model (MLSM)-- is a mixture of a Brownian motion process and a subordinator process. The subordinator process is a pure jump L\'{e}vy process. We use the mean-correction martingale measure (MCMM) method to price options and show using MCMM that our proposed pricing model is indeed arbitrage-free.

Then, following \cite{Rachev:2017}, we define a Probability Weighting Function (PWF) consistent with dynamic asset pricing theory to quantify an option trader's greed and fear disposition. The choices of PWF in  \cite{Rachev:2017} as well as in this paper guarantee that the pricing model is arbitrage-free. With the exception of \cite{ Prelec:1998},\footnote{\cite{ Prelec:1998}'s PWF maps the Gumbel distribution to another distribution. The Gumbel distribution, an infinitely divisible distribution, can be used as a model for asset pricing. Unfortunately, a pricing model with a Gumbel return distribution is overly simplistic for capturing heavy tailness and symmetry of the asset return.} all other PWFs known in the literature lead to a market model with arbitrage opportunities \cite[see][]{Rachev:2017}. To quantify an option trader's fear and greed disposition, we map the spot trader's cumulative distribution function (CDF) to another CDF corresponding to an option trader's views on the spot price for the option's underlying asset. 
In this way, we can study the fear and greed disposition of option traders using the shape of the implied PWF.
Our result shows that the PWF shape of option traders is an inverse-S-shape. This feature of PWF 
is referred to  by \cite{Tversky:1992} as \textit{diminishing sensitivity}. 

Diminishing sensitivity means that people become less sensitive to changes in probability as they move away from a reference point (see \cite{Gonzalez:1999} and \cite{Fox:1996}). In the probability domain, the two endpoints 0 and 1 serve as reference points. Thus, option traders are more sensitive to returns with a probability close to the reference points. Diminishing sensitivity results in the over-weighting of the reference points or “big losses" and “big profits." 
The PWF of option traders rises sharply near the left endpoints (events with zero probability), and steepness rising again near the right endpoint (events with probability one). This steepness indicates the fearfulness of option traders toward the market. Finally, it is worthwhile motioning that the slope of the PWF near the left endpoint, $0$, is steeper than the right endpoint and this difference strongly suggests that the significant losses are the main concern of option traders. \\
\indent There are two main contributions of this paper. First, we introduce a new L\'{e}vy process for asset returns in the form of a mixed geometric Brownian motion and subordinated L\'{e}vy process designed to describe (1) the view of the asset's spot price by spot traders and (2) the view of the asset's spot price by option traders. Second, we derive the implied PWF determining the fear and greed deposition of option traders in comparison to the spot price dynamics as viewed by spot traders.

The paper is organized as follows. After introducing the MLSM, we present the equivalent martingale measure for pricing options. We first apply the option pricing formula for a mixed subordinate normal inverse Gaussian process, and then empirically estimate the model's parameters and investigate the distribution of the log return process. We then calibrate our model parameters to the observed price of European call options based on the SPDR S$\&$P 500 ETF (SPY), followed by an investigation of the investor's fear disposition using the implied PWF we obtain.

\section*{\uppercase{option pricing for mixed subordinated L\'{e}vy process}}

In this section we derive our option pricing model where the underlying asset price is driven by a mixed subordinated L\'{e}vy process.\footnote{For a general introduction to L\textrm{\'{e}}vy processes in finance, see \cite{Sato:1999}, \cite{Bertoin:1996}, \cite{Cont:2004}, \cite{Jacod:2003}, \cite{Carr:2004}, or \cite{Schoutens:2003}.} 
\subsection*{Dynamic Asset Pricing Model}
Let $\mathcal{S}$ be a traded risky asset with price process $\mathbb{S}= (S_t,t\geq 0 )$ and log-price process $\mathbb{X}= (X_t=lnS_t,t\geq 0 )$ which is a mixed subordinated L\'{e}vy process with an added Gaussian component \citep[see][Chapter 6]{Sato:1999}. The price and log price are defined as
\begin{equation}
\label{Eq1}
S_t=S_0e^{X_t},\ t\geq 0,S_0>0
\end{equation}	

\begin{equation}
\label{Eq2}
X_t=\mu t+\varrho B_t+\sigma L_{V_t},t\ge 0,\ \ \ \mu \in R,\varrho \in R\setminus \{0\},\ \sigma \in R 
\end{equation}
where $ \mathbb{B}= (B_t,t\geq 0 )$ is a standard Brownian motion, $\mathbb{L}= (L_t,t\geq 0,L_0 = 0 )$ is a pure jump L\'{e}vy process, and $\mathbb{V}= (\ V_t,t\geq 0,V_0=0 )$ is a L\'{e}vy subordinator.\footnote{A L\textrm{\'{e}}vy subordinator is a L\textrm{\'{e}}vy processes with an increasing sample path \citep[see][Chapter 6]{Sato:1999}.} $\mathbb{E}L_1=0,\ \mathbb{E}L^2_1=1$.
Note that $\mathbb{B}$, $\mathbb{L}$, and $\mathbb{V}$ are independent stochastic bases of the natural world $ (\mathrm{\Omega },\mathcal{F},\ \mathbb{F}= ({\mathcal{F}}_t,t\ge 0 )\mathrm{,}\mathbb{P} )$. 
The trajectories of $\mathbb{L}$  and $\mathbb{V}$ are assumed to be right-continuous with left limits.

We view $\mathbb{V}$ as the $\mathcal{S}$-\textit{intrinsic} (business) time of the pure jump (the non-Gaussian, non- diffusion) part of the log return process representing the cumulative price value at time $t\geq0$ of a traded asset $\mathcal{V}$. 
We will refer to the asset $\mathcal{V}={\mathcal{V}}_{\mathcal{S}}$, as the  $\mathcal{S}$-intrinsic jump volatility. 

Parameter $\varrho \neq 0$ is the volatility of the continuous dynamics of $\mathbb{X}$, and $\sigma$ is the volatility of the pure jump of the subordinated process $L_{V_t}$.

\subsection*{Equivalent Martingale Measure}
Let $\mathcal{B}$ be a riskless asset with price $b_t=e^{rt},t\ge 0$, where $r\ge 0$ is the riskless rate. For the pricing of financial derivatives, we search for an equivalent martingale measure (EMM) $\mathbb{Q}$ of $\mathbb{P}$ on $ (\mathrm{\Omega },\mathcal{F},\ \mathbb{F}= ({\mathcal{F}}_t,t\ge 0 )\mathrm{,}\mathbb{Q} )$. The discounted price process $Z_t=\frac{S_t}{b_t}$ is a martingale.\footnote{ See \citet[Chapter 6]{Duffie:2001}, and \citet[Section 2.5]{Schoutens:2003}.}

The market $ (\mathcal{S},\mathcal{B} )$ is incomplete and the solution of EMM is not unique. It is generally accepted that the MCMM is sufficiently flexible for calibrating market data.\footnote{ See \citet[Chapters 6 and 7]{Schoutens:2003}. It is tempting to find a EMM using the Esscher transform (see \cite{Esscher:1932}, \cite{Gerber:1994}, \cite{Salhi:2017}), as in this case we can set $\varrho =0$. However, with $\varrho =0$, the Esscher transform method requires finding a unique solution $h^*$ of the equation: $r=\mu +K_{V_1} ( K_{ L_1} ( (u+1 )\sigma  ) )-K_{V_1} (K_{L_1} (h\sigma))$,  where $K_{L_1 } (u )=ln\mathbb{E}e^{uX_1},K_{L_1} (u )=ln\mathbb{E}e^{uL_1}$ and  $K_{V_1} (u )=ln\mathbb{E}e^{uV_1}$, $u\in R$, are the cumulant-generating functions for $\mathbb{X}$, $\mathbb{L}$ and $\mathbb{V}$. In the general setting of \eqref{Eq2}, this is an impossible task.} Thus, we choose MCMM as the risk-neutral probability space, $\mathbb{Q}$. 
\cite{Yao:2011} demonstrated that $\mathbb{Q}$ obtained by the MCMM is equivalent to  $\mathbb{P}$ if and only if the Gaussian part in the L\textrm{\'{e}}vy-Khintchine formula for the characteristic function of $\mathbb{X}$ is non-zero. If $\mathbb{X}$ is a pure jump L\textrm{\'{e}}vy process, the MCMM $\mathbb{Q}$ is not equivalent to $\mathbb{P}$. However, because the European call option pricing formula under  $\mathbb{Q}$ is still arbitrage free, the price dynamics of $\mathcal{S}$ on $\mathbb{Q}$ is given by
\begin{equation}
\label{Eq3}
S^{(\mathbb{Q})}_t=S_0\frac{b_t}{M_{X_t} (1)}e^{X_t}=S_0e^{(r-K_{X_1}(1))t+X_t},t \geq 0
\end{equation}
where the moment-generating functions (MGF) $M^{ (\mathbb{X} )}_t$ and the cumulant-generating function (CGF) $K^{ (\mathbb{X} )}_t$ of the L\'{e}vy process $\mathbb{X}\ $ are
\begin{equation}
M_{ X_t } (u )=\mathbb{E}e^{uX_t}={(M_{X_1}(u))}^t,u\ge 0
\end{equation}	
\begin{equation}
K_{ X_t}(u)=lnM_{X_t}(u),\ u\ge 0,\ t\ge 0
\end{equation}		

Similarly, let $M_{ L_t}$ and $M_{V_t}$, $u\in R$, $t\ge 0$ be the MGFs of $\mathbb{L}$ and $\mathbb{V}$, respectively. And let $K_{L_t}$ and $K^{V_t}$ be the CGFs of  $\mathbb{L}$ and $\mathbb{V}$, respectively.  We then have

\begin{equation}
\label{Eq4}
K_{ X_1} (1)=\mu +\frac{{\varrho }^2}{2}+K_{ V_1} (K_{L_1} (\sigma))\ <\infty 
\end{equation}

\subsection*{Option Pricing Model}
Let $\mathcal{C}$ be a European call contract with underlying risky asset $\mathcal{S}$, maturity $T>0$, and strike $K>0.$ Then the price of $\mathcal{C}$ at $t=0,$ is given by
\begin{equation}
\label{Eq5}
C (S_0,r,K,T )=e^{-rT}{\mathbb{E}}^{\mathbb{Q}}{\mathrm{max}  (S^{ (\mathbb{Q} )}_T-K,0 )}
\end{equation}

Carr and Madan (1998)
showed that if $a>0$, which leads to ${\mathbb{E}}^{\mathbb{Q}}{ (S^{ (\mathbb{Q} )}_T )}^{a}<\infty $, then
\begin{equation}
\label{Eq6}
C (S_0,r,K,T )=\frac{e^{-rT-ak}}{\pi }\int^{\infty }_0{e^{-ivk}}\frac{{\varphi }_{lnS^{ (\mathbb{Q} )}_T} (v-i(a+1) )}{{a}^2+a -v^2+i (2a +2 )v}dv
\end{equation}
where $k=lnK$ and ${\varphi }_{lnS^{ (\mathbb{Q} )}_t} (v )={\mathbb{E}}^{ (\mathbb{Q} )}e^{ivlnS^{ (Q )}_t}$ is the characteristic function (ch.f.) of the log-price process  ${\mathbb{L}\mathbb{S}}^{ (\mathbb{Q} )}= (lnS^{ (\mathbb{Q} )}_t,t\ge 0 )$.  

From \eqref{Eq3} and \eqref{Eq4} the ch.f. ${\varphi }_{lnS^{ (\mathbb{Q} )}_t}$ of the log-price process  ${\mathbb{L}\mathbb{S}}^{ (\mathbb{Q} )}$ is given by 
\begin{equation}
\label{Eq7}
\begin{split}
{\varphi }_{lnS^{ (\mathbb{Q} )}_t} (v )&=S^{iv}_0e^{iv (r-K_{X_1} (1) )t}{\varphi }_{X_t}(v)
\\
&=S^{iv}_0{\mathrm{exp}  \{ [iv (r-K_{X_1}(1))+{\psi }_{X_t} (v) ]t\}}
\end{split}
\end{equation}	
where ${\varphi }_{ X_t} (v)=\mathbb{E}e^{ivX_t}$ is the ch.f. of $\mathbb{X}$ and  ${\psi }_{ X_t} (v )=ln{\varphi }_{ X_t}(v)$ is the characteristic exponent of $\mathbb{X}$. 

Similarly, the characteristic functions and corresponding characteristic exponents for  $\mathbb{L}$ and $\mathbb{V}$ are ${\varphi }_{ L_t}$, ${\psi }_{L_t}$, ${\varphi }_{ V_t}$, and ${\psi}_{ V_t}$. And the domain of those functions and exponents are complex planes.
From \cite{Sato:1999}, the exponential moment conditions guaranteeing that  ${\psi }_{L_t} (v ),\ v\in \mathbb{C}$ and  ${\psi }_{V_t} (v ),\ v\in \mathbb{C}\mathrm{,}$ are well defined. Then, we have

\begin{equation}
\label{Eq8}
{\psi }_{X_t}(v)=iv\mu -\frac{{\varrho }^2}{2}v^2+{\psi}_{V_t} (-i{\psi }_{ L_t} (v\sigma)),\,v\in \mathbb{C}.
\end{equation}
Thus, we derive the call option price $C (S_0,r,K,T )\ $ in \eqref{Eq5} using \eqref{Eq6}, \eqref{Eq7}, and \eqref{Eq8}.

\section*{\uppercase{option pricing for mixed subordinated normal inverse Gaussian process}}

In this section, we apply the European call option pricing formula \eqref{Eq6} where $\mathbb{L}$ is the Normal Inverse Gaussian (NIG) L\'{e}vy process\footnote{ See \cite{Barndorff:1994}, \cite{Eriksson:2009}, and \citet[][Section 5.3.8]{Schoutens:2003}.}  and $\mathbb{V}$ is the Inverse Gaussian (IG) L\'{e}vy subordinator.\footnote{ See \citet[][Chapter 12]{Barndorff:2015}  and \citet[][Section 5.3.2]{Schoutens:2003}.} Then, the CGF $K_{ L_1}$ of the NIG process $\mathbb{L}$ has the following parametric form:
\begin{equation}
\label{Eq9}
K_{L_1}(u)=mu+d (\sqrt{\alpha^2-\beta^2}-\sqrt{\alpha^2-{ (\beta+u )}^2} ),u\in  (-\alpha-\beta,\alpha-\beta )
\end{equation}
where $m\in R$ is the location parameter, $\alpha>0$ is the tail-heaviness parameter, $\beta\in R$ ($ \lvert \beta \rvert < \alpha$) is the asymmetry parameter, and $d$ is the scale parameter. Then the CGF $K{V_1}$ of the IG subordinator $\mathbb{V}$ is given by
\begin{equation}
\label{Eq10}
K_{V_1}(u)=\frac{\ell }{h} (1-\sqrt{1-\frac{2h^2u}{\ell }} ),u\in  (0,\frac{\ell }{2h^2} ),
\end{equation}	
where $k >0$ is the mean of $V_1$ and $\ell>0$ is the shape parameter for the IG distribution.

\subsection*{Characterization of the distributional  law of process $\mathbb{X}$ with log-price  }

We now study the ch.f and the cumulant of 	 
$ X_t=\mu t+\varrho B_t+\sigma L_{V_t}$, $t\ge 0$, $\mu \in R$, $\varrho \in R\setminus  \{0 \}$, and $\sigma \in R$. The ch.f of ${X_1}$ has the form

\begin{equation}
\label{Eq11}
\begin{array}{c}
\varphi_{X_1}(v)=e^{iv \mu - \frac{1}{2} \rho^2 v^2+\frac{l}{h}
	[1-\sqrt{1-\frac{2h^2}{l} [d  [ \sqrt{\alpha^2-\beta^2}-\sqrt{\alpha^2- ( \beta+\sigma iv )^2 }  ] 
		+iv m \sigma	 ] 
	}
	] 
}, v\in \mathbb{C}
\end{array}
\end{equation}

The MGF of $X_1$, $M_{X_1}(u)$, is obtained with $v=\frac{u}{i}$
\begin{equation}
\label{Eq12}
\begin{array}{c}
M_{X_1}(u)=e^{u \mu + \frac{1}{2} \rho^2 u^2+\frac{l}{h}
	[1-\sqrt{1-\frac{2h^2}{l} [d  [ \sqrt{\alpha^2-\beta^2}-\sqrt{\alpha^2- ( \beta+\sigma u )^2 }  ] 
		+u m \sigma	 ] }] }.
\end{array}
\end{equation}
with the constraints 
\begin{equation}
\label{MGF_condition}
0<u< (\frac{\alpha-\beta}{\sigma}  )
\end{equation}
\begin{equation}
u  ( m \sigma )  +d  ( \sqrt{\alpha^2-\beta^2}-\sqrt{\alpha^2- ( \beta+\sigma u )^2 }  )   <\frac{l}{2h^2}
\end{equation}

In this case, $X_1$ has a finite exponential moment for any $u$ in \eqref{MGF_condition}. 
From the representation of the MGF, we can determine all four moments of $X_1$. To find the four central moments of $X_1$, we use the CGF
${\mathrm{K}}_{\mathrm{X_1}} (v )=ln{\ M}_{\mathrm{X_1}} (v )$, and the cumulants ${\kappa }_n\mathrm{=}{ [\frac{{\partial }^n}{\partial u^n}K_{X(1)} (u ) ]}_{u=0},\ n=1,2,3,4$. 
The CGF is 
\begin{equation}
\label{Eq13}
\begin{split}
K_{ X_1}(u)=u \mu + \frac{1}{2} \rho^2 u^2+\frac{l}{h} [1-\sqrt{1-\frac{2h^2}{l} [d[ \sqrt{\alpha^2-\beta^2}-\sqrt{\alpha^2- ( \beta+\sigma u )^2 }]+u m \sigma]}]
\end{split}
\end{equation}

Then, we have
\begin{align*}
\mathbb{E}(X_1)&={\mathrm{\kappa }}_{1}
\\
Var (X_1 )&={\mathrm{\kappa }}_{2}
\\
{Skewness} (X_1)&=\frac{\mathbb{E}{ [X_1-\mathbb{E}X_1 ]}^3}{{ [var(X_1) ]}^{\frac{3}{2}}}=\frac{{\kappa }_{3}}{{ ({\kappa }_{2} )}^{\frac{3}{2}}}
\\
{Excess Kurtosis}  (X_1 )&=\frac{\mathbb{E}{ [X_1-\mathbb{E}X_1 ]}^4}{{ [var(X_1) ]}^2}-3=\frac{{\kappa }_{4}}{{ ({\kappa }_{2} )}^2}.
\end{align*}

More specifically,, the mean of $X_1$ is given by
\begin{equation}
\label{Eq14}
E(X_1)=\mu + h (  m\sigma + \frac{\beta d \sigma}{\sqrt{\alpha^2 - \beta^2}} )
\end{equation}	
For the variance of $X_1$ we have
\begin{equation}
\label{Eq15}
Var(X_1)=h ( \frac{ d\sigma^2} { \sqrt{ \alpha^2-\beta^2}}
+\frac{\beta^2 d \sigma^2}{ ( \alpha^2-\beta^2 )^{\frac{3}{2}}} ) 
+\rho^2
+\frac{h^3  ( m \sigma +\frac{ \beta d \sigma}{\sqrt{\alpha^2-\beta^2}}  )^2}{l}
\end{equation} 
The skewness and kurtosis are obtained by applying the same method, and thus are omitted. 

\subsection*{Option pricing with log-price process $\protect\mathbb{X}$ }
Carr and Madan (1998) developed an explicit pricing method for vanilla options when the characteristics function of the log-price process under the risk-neutral world is known. If we know the ch.f. of  $lnS^{ (Q )}$, we can calculate the price of a call option by applying \eqref{Eq6}. From \eqref{Eq3} and \eqref{Eq4}, we can derive the ch.f. of the log-price process ${\mathbb{L}\mathbb{S}}^{ (\mathbb{Q} )}= (lnS^{ (\mathbb{Q} )}_t,t\ge 0 )$ as follows 
\begin{equation}
\label{Eq16}
\begin{split}
{\varphi }_{lnS^{ (\mathbb{Q} )}_t} (v )&={\mathbb{E}}^{ (\mathbb{Q} )}e^{ivlnS^{ (Q )}_t}
\\	
&=S^{iv}_0e^{ivrt-\frac{1}{2}vt\rho^2(i+v)-P_1-P_2}
\end{split}
\end{equation}
where 
\begin{align*}	
P_1&=\frac{l}{h}t(1+iv-iv\sqrt{1-\frac{2h^2}{l}[d[\sqrt{\alpha^2-\beta^2}-\sqrt{\alpha^2-(\beta+\sigma)^2}]+m\sigma]}
\\
P_2&=\sqrt{1+d[\sqrt{\alpha^2-\beta^2}-\sqrt{\alpha^2-(\beta+\sigma iv)^2}]+ivm\sigma]}
\end{align*}

To determine the price of a call option, we substitute \eqref{Eq16} into \eqref{Eq5} and perform the
required integration. We use the fast Fourier transformation (FFT) to estimate the call option price in \eqref{Eq6} with strike $K$, time to maturity $T$, and risk-free rate $r$ at time $0$.

\section*{\uppercase{numerical example}}
In this section, we apply the method introduced in the previous section. We use the historical data of the S$\&$P 500 index\footnote {See \url{https://us.spdrs.com/en/etf/spdr-sp-500-etf-SPY}} and CBOE volatility index (VIX) \footnote{VIX is an index created by CBOE, representing 30-day implied volatility calculated by S\&P500 options, see {http://www.cboe.com/vix.}} to estimate the model parameters for spot traders, while using the call option prices for the SPRD S\&P 500 ETF (SPY) \footnote{https://finance.yahoo.com/quote/SPY/options?p=SPY} as the dataset to estimate the model parameters for option traders.

\subsection* {Fitting the spot market data}

In this subsection, we apply the models we proposed earlier to estimate the returns of a broad-based market index (the S$\&$P 500) whose return is measured by the return of an exchange-traded fund, SPY. We use market indices by the pair  $ (X_t,V_t )$, $t\geq 0$ where $(1)\ X_t$, $t\geq$ as a stochastic model for the log-return of SPY  index,  and $(2)\ V (t )$, $t\ge0$ as the cumulative VIX (i.e., $V(t)$ represents the cumulative value of VIX in $ [0,t ]$), where $ (a)\ X_t$, $t\geq$ is a stochastic model for the log-return of the SPY and $ (b)\ V (t )$, $t\geq0$ is the cumulative VIX (i.e., $V (t )$).

We then fit the IG distribution to the daily VIX data and evaluate the density using PP-plot, goodness-of-fit test, and the probability integral transforms (PIT) test. The mean $(h)$ and shape $(l)$ are $0.192548$ and $1.49156$ respectively, fitted by maximum likelihood method on daily VIX index data from January 1993 to the end of March 2019.  

Exhibit 1 shows the fitted results of the empirical CDF and theoretical CDF and the PP-plot of IG. Our estimated model performs well with respect to the CDF and the empirical CDF fitting process. Moreover, the apparent linearity of the PP-plot shows that the corresponding distributions are well-fitted. The Kolmogorov--Smirnov test gives a P--value$ ( \simeq0.062  )$, meaning that it fails to reject the null hypothesis that our model is sufficient to describe the data.

\begin{figure}
	\begin{subfigure}{200pt}
		\includegraphics[width = 200pt]{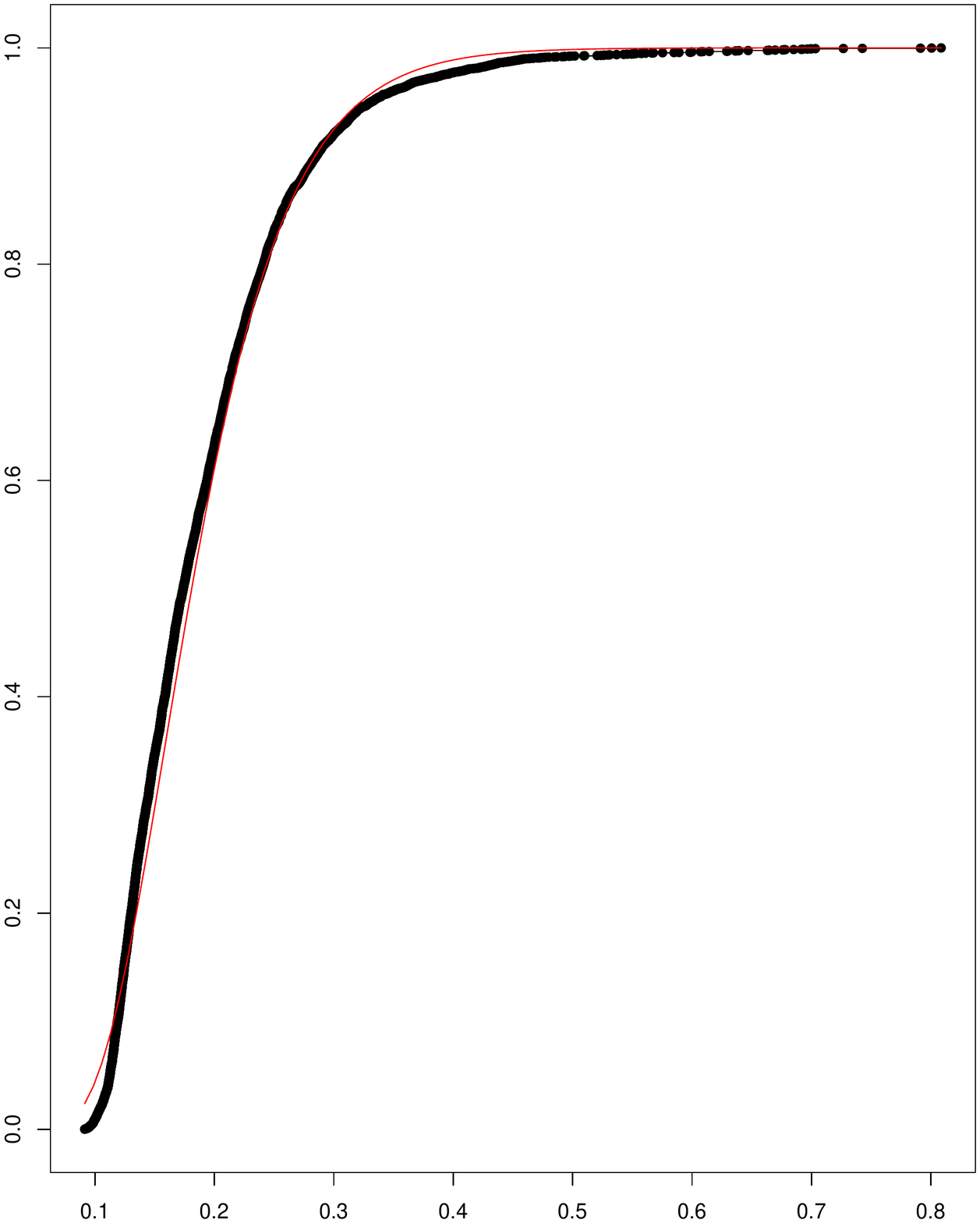}
		\caption*{Empirical and theoretical CDFs}    
	\end{subfigure} 
	\begin{subfigure}{250pt} 
		\includegraphics[width=200pt]{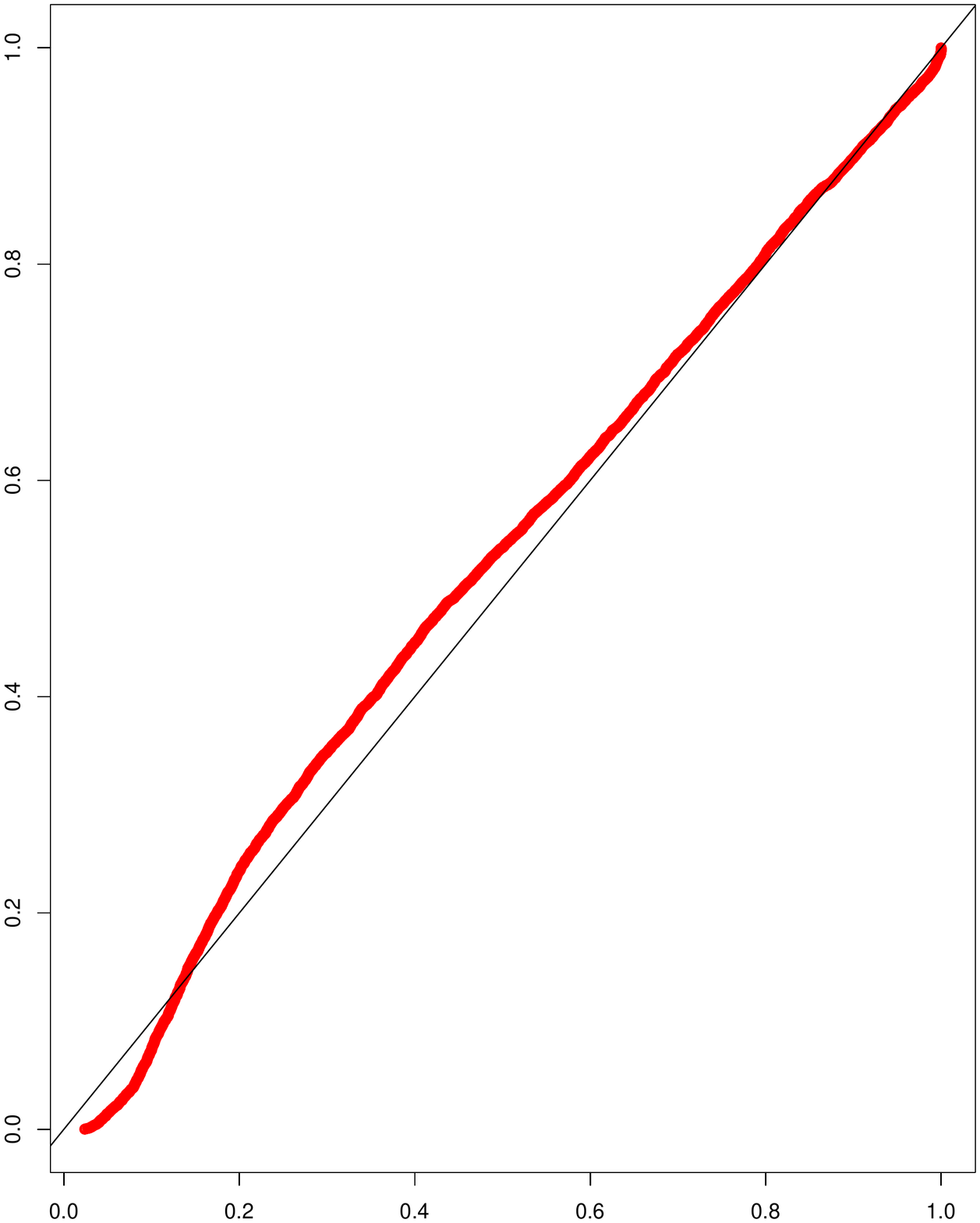}
		\caption{IG PP-plot of VIX data}
	\end{subfigure}
	\caption*{Exhibit 1: The IG fitting results}
	\label{the_IG_fitting}
\end{figure}

We then investigate the distribution of 
\begin{equation*}
X_t=\mu t+\varrho B_t+\sigma L_{V_t}, t\ge 0, \mu \in R, \varrho \in R\setminus  \{0 \}, \sigma \in R
\end{equation*}
as the stochastic model for the SPY log-return index by fitting the distribution derived from the ch.f. of $X_t$ to the data. Among the $10$ parameters of stochastic process $X_t$, for two of them $( l,h )$, the parameter of the intrinsic time-change process $V_t$, is estimated by fitting the IG distribution to the VIX data. Instead of using the maximum likelihood method for the other parameters, we apply model fitting via the empirical characteristic function (ECF) \citep[see][]{Yu:2003} to estimate the model parameters. 
Notice that the probability density function (pdf) is the FFT of the ch.f.. The existence of a one-to-one correspondence between the CDF and the ch.f. makes inference and estimation using the ECF method as efficient as the maximum likelihood method. To estimate the model parameter, we minimized
\begin{equation}
\label{emprical_ch_minimization}
h (r,x,\theta  )=\int_{-\infty}^{\infty} ( \frac{1}{n}\sum_{i=1}^{n} e^{i\theta x_i} -C ( r,\theta ) )  ^2 dr
\end{equation}  
where $C(r,\theta)$ is the ch.f. of $X_t$ given by \eqref{Eq11}. The database covers the period from January 1993 to March 2019, a total 6,591 observations collected from Yahoo Finance. 

The initial values are obtained using the method of moments estimation and by making
instructed guesses. For any initial value, we estimated the model parameters and consider
the model as a good candidate to fit the data.
We implemented the FFT to calculate both the pdf and the corresponding likelihood values. The best model to fit and explain the observed data is chosen as the one with the largest likelihood value. 

The estimated parameters are summarized in Exhibit  2. The model density estimates corresponding to the empirical density of the daily log-return SPY index are plotted in Exhibit 3. The Exhibit reveals that our estimated model offers a good match between the pdf and the empirical density of the data. In our estimation  $E(L_1)=m+\frac{d\beta}{\sqrt(\alpha^2-\beta^2)}\approx0$ and $Var(L_1)=\frac{d\alpha^2}{ ( \sqrt(\alpha^2-\beta^2) ) ^3}\approx 1$.	

\begin{table}[htb]
	\caption*{ Exhibit 2: The estimated parameters of  the distribution fitted to daily SPDR S\&P 500  log-returns.}
	\label{tab:Table_Option_trader}
	\begin{tabularx}{\textwidth}{c *{8}{Y}}
		\toprule[1pt]
		{$\mu$}&{$m$}&{$\alpha$}&{$\beta$}&{$d$}&{$\rho$}&{$\sigma$}\\
		\hline
		0.00002&-0.00018&310.8&1.19&0.007&0.0011&2.199\\
		\bottomrule[1pt]
	\end{tabularx}
\end{table}
\begin{figure}[htb]
	\centering
	\includegraphics[width=1.0\textwidth]{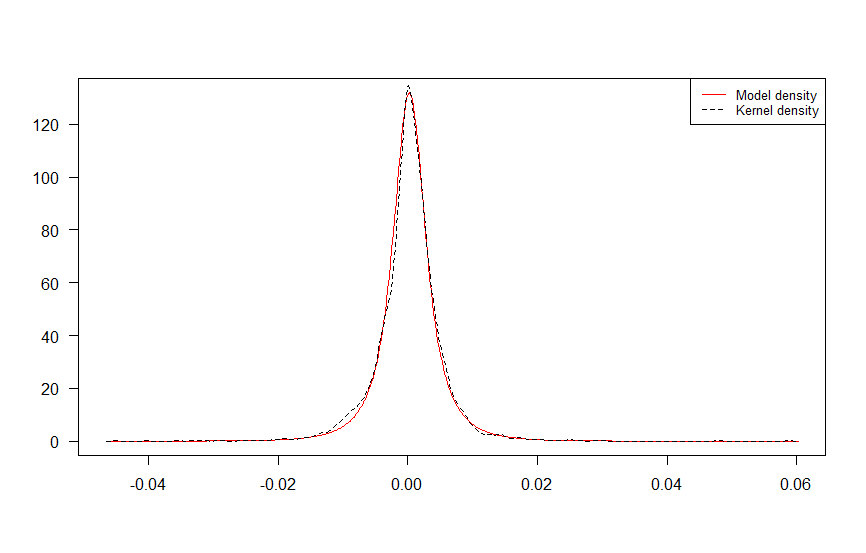}
	\caption*{Exhibit 3: The density of log-return SPDR S\&P 500  via the kernel density.}
	\label{figure:SPY_fit}
\end{figure}

\subsection*{Calibration of the spot market data}
We now apply our mixed subordinated L\'{e}vy process model to price a European vanilla option on the SPY index.  First, we calibrate the parameters of the model's risk-neutral probability measure. The calibration is performed by implementing the ``Inverse of the Modified Call Price" methods introduced by \cite{Carr:1998}. 

The data we use for call option prices are from Yahoo Finance for 08/29/2019 with different expiration dates and strike prices. The expiration date varies from 08/30/2019 to 12/17/2021, and the strike price varies from \$25 to \$430 among 2,440 different call option contracts. As the underlying of the call option, the SPY index price is \$292.58 on 08/29/2019. We use the 10-year Treasury yield curve rate\footnote{https://www.treasury.gov/resource-center/data-chart-center/interest-rates/Pages/TextView.aspx?data=yieldYear\&year=2019} on 08/29/2019 as the risk-free rate $r$, here $r=0.015$. Following \cite{Schoutens:2003}, we set $a=0.75$ and calibrate parameters from call option prices by \eqref{Eq6}. The estimated parameters of the best model are reported in Exhibit 4.

\begin{table}[htb]
	\caption*{Exhibit 4: The calibrated parameters fitted call option prices on 08/29/2019.}
	\label{TableStock_trader}
	\begin{tabularx}{\textwidth}{c *{7}{Y}}
		\toprule[1pt] 
		{$m$}&{$\alpha$}&{$\beta$}&{$d$}&{$\rho$}&{$\sigma$}\\
		\hline
		-0.4&241&1.2&5&0.05&2.13\\
		\bottomrule[1pt]
	\end{tabularx}
\end{table}

We use the inverse FFT and nonlinear least-squares minimization strategy to calibrate the parameters. As shown in Exhibit 4, the calibrated parameters have similar values to those reported in Exhibit 2, which is from the spot SPY and VIX. Note that the same method can be applied to put options. Since the model parameters are estimated from call option data, the model is the asset log-return model observed by option traders.

\section*{\uppercase{implied Probability Weighting function}}

The general framework of behavioral finance provides an alternative view of the mixed subordinated price process \citep[see][]{Barberis:2003}. \cite{Tversky:1992} introduced the Cumulative Prospect Theory (CPT). According to this theory, positive and negative returns on financial assets are treated differently due to the general fear disposition of investors.

To quantify an investor's fear disposition, \cite{Tversky:1992} and \cite{Prelec:1998} introduced a PWF,
$w^{ (\mathcal{R},\mathcal{S} )}: [0,1 ]\to  [0,1 ]$, 
transforming the asset return distribution given by
\begin{equation*}
F_{\mathcal{R}} (x )\mathrm{=}\mathbb{P} (\mathcal{R}\le x ),x\in R
\end{equation*}
to a new one given by
\begin{equation*}
F_{\mathcal{S}} (x )\mathrm{=}\mathbb{P} (\mathcal{S}\le x )=w^{ (\mathcal{R},\mathcal{S} )} (F_R (x ) ),x\in R
\end{equation*} 
corresponding to an option trader's views. 

\cite{Tversky:1992} introduced the following PWF
\begin{equation}
\label{PWT}
w^{ (\mathcal{R},\mathcal{S};TK )} (u )=\frac{u^{\gamma }}{{ [u^{\gamma }+{ (1-u )}^{\gamma } ]}^{\frac{1}{\gamma }}},\,\,u\in  (0,1 ),\,\, \gamma \in  [0,1].
\end{equation} 
This PWF corresponding to $F_{\mathcal{S}} (x )$  requires an infinitely divisible distribution of the asset return. If not, it would lead to arbitrage opportunities in behavioral asset pricing models. \cite{Rachev:2017} studied the general form of PWF consistent with dynamic asset pricing theory. They treated $\mathcal{R}=M_t$, $t\ge0$ as the asset price dynamics before introducing the views of investors,  where $\mathcal{R}=M_t$, a single subordinated log-price process, is given by 

\begin{equation}
\label{single_subordinated}
M_t=\mu t+\gamma U (t )+\sigma B_{U(t)}\, ,\,\,t\ge 0,\,\,\mu \in R,\,\,\gamma \in R,\,\,\sigma >0
\end{equation} 

The investor’s fear can be taken into account by introducing a new log-price process with a second ``behavioral" subordinator \citep[see][]{Shirvani:2019}. In our work, the investor's fear is incorporated into the BSM asset return model by  introducing a pure jump L\'{e}vy process $L_t$ with $\mathbb{E}L_1=0,\ \mathbb{E}L^2_1=1$. The new mixed L\'{e}vy process is

\begin{equation}
X_t=\mu t+\varrho B_t+\sigma L_{t},\,\,t\ge 0,\  \mu \in R,\,\varrho \in R\diagdown  \{0 \},\ \sigma \in R
\label{investor_view_equation}
\end{equation}
The ch.f., $\varphi_{X_1}(v)$ has the form
\begin{equation}
\label{Chf_investor}
\begin{array}{c}
\varphi_{X_1} (v )=e^{iv \mu - \frac{1}{2} \rho^2 v^2+iv m \sigma +d  [ \sqrt{\alpha^2-\beta^2}-\sqrt{\alpha^2- ( \beta+\sigma iv )^2 }  ] 
}, v\in \mathbb{C}.
\end{array}
\end{equation}
The MGF of $X_1$, $M_{X_1} (u )$, is obtained by setting  $v=\frac{u}{i}$
\begin{equation}
\label{MGF_trade}
\begin{array}{c}
M_{X_1} (u )=e^{u \mu + \frac{1}{2} \rho^2 u^2+u m \sigma +d  [ \sqrt{\alpha^2-\beta^2}-\sqrt{\alpha^2- ( \beta+\sigma u )^2 }  ] 
}, u\in  (0,\frac{\alpha-\beta}{\sigma} )
\end{array}
\end{equation}

The corresponding PWF, $w^{ (\mathcal{R},\mathcal{S} )}: [0,1 ]\to  [0,1 ]$,  is defined by
\begin{equation*}
w^{ (\mathcal{R},\mathcal{S} )} (u )=F_{\mathcal{S}} (F^{inv}_{\mathcal{R}} (u ) )
\end{equation*}
where $F^{inv}_{\mathcal{R}} (u )={\mathrm{min}  \{x:F_{\mathcal{R}} (x )>u \}\ }$ is the inverse function of $F_{\mathcal{R}} (x )$ \citep[see][]{Rachev:2017}.

This PWF $w^{ (\mathcal{R},\mathcal{S} )}$ represents the views of the option trader on the spot market model. These views about the market are different from those of a spot trader. In general, option traders are more ``fearful" than spot traders due to the non-linearity of the risk factors they face. 

To study whether option traders are greedy or fearful, we need to calculate  $w^{ (\mathcal{R},\mathcal{S} )}$ and focus on the shape of PWF. To do so, we calculate the PWF of option traders by transforming the spot trader's distribution to the corresponding option trader's distribution where the asset log-return process follows \eqref{investor_view_equation}. We take $\mathcal{R}=X_t$, $t\ge0$ as the dynamics of the current log-price return observed by spot traders if the parameters of $X_t$, $t\ge0$ are estimated from the spot market or the natural world. Moreover, we consider $\mathcal{S}=X^{risk-neutral}_t$ as the dynamics of the log-price return observed by option traders where  $X^{risk-neutral}_t$ is
\begin{equation*}
X^{risk-neutral}_t=X^{ (\mathbb{Q} )}_{t}=r t+\varrho B_t+\sigma^{\mathbb{Q}} L^{\mathbb{Q}}_{t}\,,\, \varrho \in R\setminus  \{0 \}, \sigma^{\mathbb{Q}} \in R
\end{equation*}
where $\varrho$ is estimated from the spot prices of the underlying asset. The remaining parameters for the distribution of $X^{risk-neutral}_t$ are calibrated from the risk-neutral world.

To estimate the parameters in $\mathcal{R}=X_t$, where $X_t$ represents the dynamics of the log-price return observed by spot traders, we applied the ch.f. method to daily log-returns (based to closing prices) of the SPY from January 1993 to March 2019. The model's estimated parameters are summarized in Exhibit  5.  We implemented the FFT to calculate the CDF of the model.  The result, plotted in Exhibit 6, shows that our estimated model provides a good match between the CDF and the CDF of the data.
\begin{table}[htb]
	\begin{center}
		\label{taboptin_trade}
		\caption*{Exhibit 5: The estimated parameters of the distribution of spot traders fitted to daily SPDR S\&P 500  log-returns}
		\begin{tabularx}{\textwidth}{c *{7}{Y}}
			\toprule[1pt]
			{$m$}&{$\alpha$}&{$\beta$}&{$d$}&{$\rho$}&{$\mu$}&{$\sigma$}\\
			\hline
			0.00039&176.8&3.45&0.0025&0.0011&-0.00008&1.399\\
			\bottomrule[1pt]
		\end{tabularx}
	\end{center}
\end{table}

\begin{figure}[t!]
	\label{figureTrader_fit}
	\centering
	\includegraphics[width=1.0\textwidth]{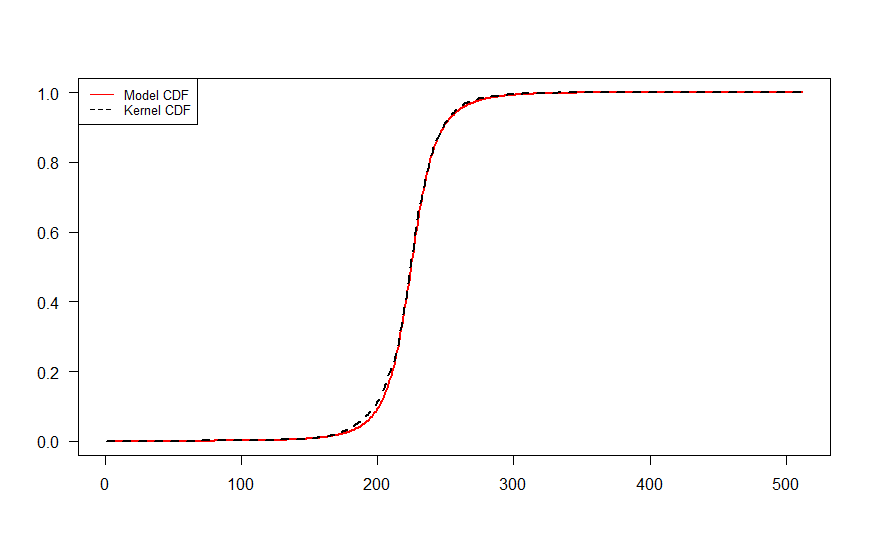}
	\caption*{Exhibit 6: The CDF of the spot trader model via the kernel density.}
\end{figure}

We calibrate the parameters of  $\mathcal{S}=X^{risk-neutral}_t$ in the risk-neutral probability space using the ``Inverse of the Modified Call Price" methods (\cite{Carr:1998}).  Let $\mathcal{S}$ be a traded risky asset with price process
\begin{equation*} 
S_t=S_0 e^{X_t},\ t\ge 0,S_0>0
\end{equation*}
where the log-price process $\mathbb{X}= (X_t=lnS_t,t\ge 0 )$ is a mixed L\'{e}vy process:
\begin{equation*}
X_t=\mu t+\varrho B_t+\sigma L_{t},\,\,t\ge 0,\  \mu \in R,\,\varrho \in R\setminus  \{0 \},\ \sigma \in R
\end{equation*}
Since $X_t$ is a pure jump L\textrm{\'{e}}vy process, the MCMM $\mathbb{Q}$ is not equivalent to $\mathbb{P}$, while the European call option pricing formula under  $\mathbb{Q}$ is still arbitrage free. 
The ch.f. for $X_t$, with $ v\in \mathbb{C}$ , is given by
\begin{equation}
\label{Chf_option_trader}
{\varphi }_{lnS^{ (\mathbb{Q} )}_t} (v )=S^{iv}_0e^{ivrt- \frac{1}{2} vt\rho^2  (i+v  )  
	-ivtd [ \sqrt{\alpha^2-\beta^2}-\sqrt{\alpha^2- ( \beta+\sigma )^2 }  ] 			
	+td  [ \sqrt{\alpha^2-\beta^2}-\sqrt{\alpha^2- ( \beta+\sigma iv )^2 }  ] 		
}
\end{equation}

To calibrate our model parameters, we use the same dataset of call option prices in the previous section. 
The 10-year Treasury yield curve rate is regarded as the risk-free rate $r$. According to \cite{Schoutens:2003}, we set $a=0.75$ and calibrate parameters based on call option prices by \eqref{Eq6} with the same methods mentioned in the previous section and construct the CDF of option traders.

Using the CDFs of $\mathcal{S}$ and $\mathcal{R}$, we numerically computed the corresponding PWF, $w^{ (\mathcal{R},\mathcal{S} )}$.  
\cite{Gonzalez:1999} discuss two features of PWF: \textit{Diminishing sensitivity and discriminability} and \textit{attractiveness}. They  interpreted the discriminability as the degree of curvature of the PWF and attractiveness as the elevation of the PWF.  \cite{Tversky:1992} presented a psychological definition for diminishing sensitivity as: people are less sensitive to change in probability as they move from reference points. Zero and one refer to reference points in the probability domain. The plotted PWF in Exhibit 7 shows diminishing sensitivity of option traders. As shown in Exhibit 7, the PWF has an inverse-S-shape, first concave and then convex. 
The plot falls sharply near the probability value $0.17$ and rises steeply near the point $0.95$ to $1$. The PWF varies slightly in interval $ (0.1,0.9)$, indicating that option traders overestimate the probability of values that are not close to  reference points. In other words, option traders overweight the probability of big losses and underweight the probability of big profits. The falling near to zero and rising near the endpoint (concave, then convex) of the PWF, represent option trader's fear of a  big jump in the market, especially for big losses. That is, option traders tend to be more fearful than spot traders. The second feature of the PWF discussed by \cite{Gonzalez:1999} is not related to the shape and curvature of the PWF and therefore beyond the scope of this paper.

\begin{figure}[t!]
	\centering
	\includegraphics[width=.50\textwidth]{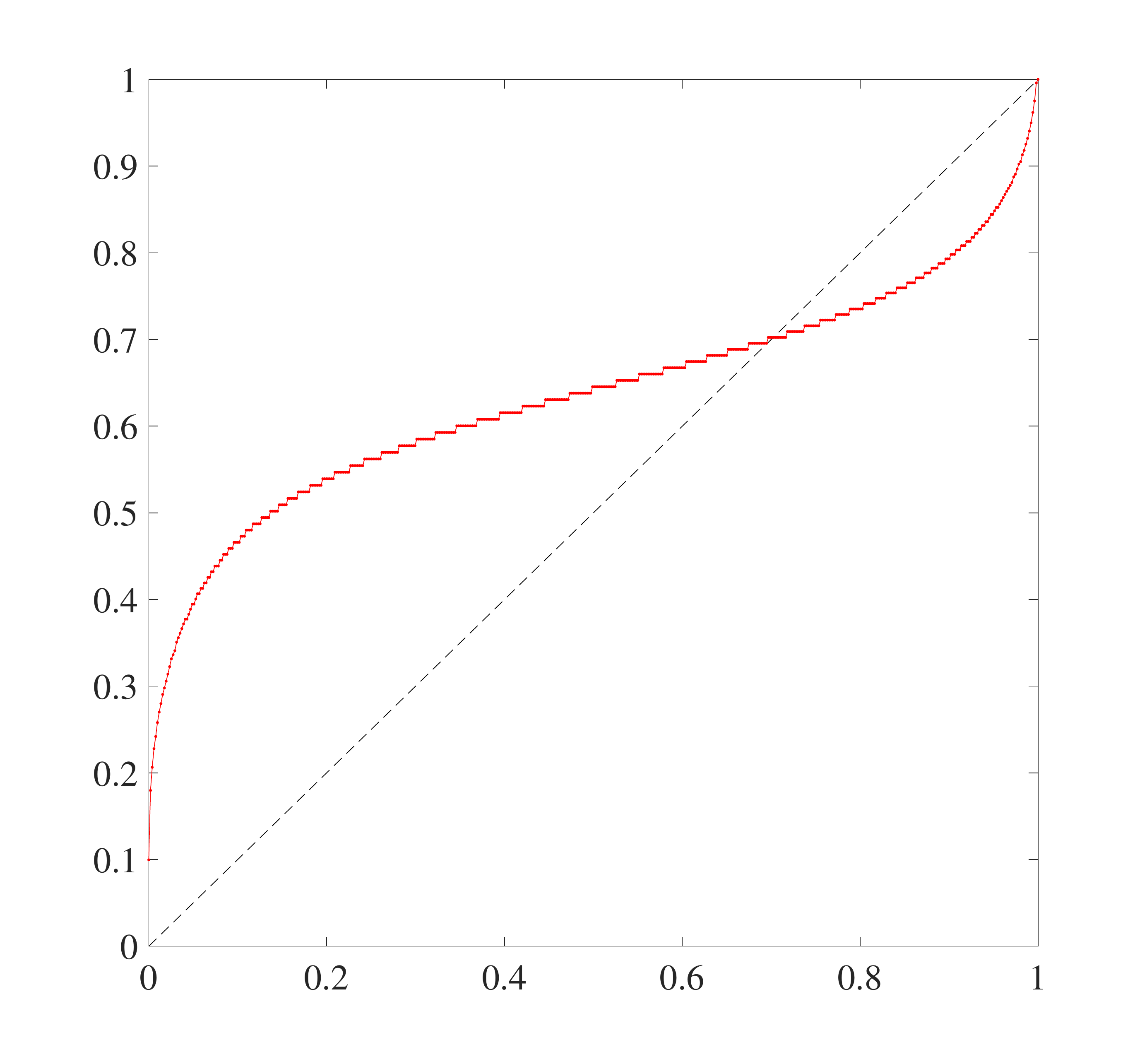}
	\caption*{Exhibit 7: The probability weighting function of option trader.}
	\label{figurePWF}
\end{figure}

\section*{\uppercase{conclusion}}

In this paper, we develop a more realistic asset pricing model by mixing the BSM asset return process with a single L\'{e}vy subordinated process, through which
we are able to incorporate the behavior and sentiment of investors in a log-return pricing model. 
Then we present the arbitrage-free equivalent market measure.
We apply the European call option pricing formula where the subordinated process is a Normal Inverse Gaussian L\'{e}vy process. The model parameters are calibrated using the SPY index. 
The investor’s fear disposition is evaluated by the PWF. We reviewed the shape of the weighting function in terms of discriminability. The PWF shape of option traders starts out as concave and then becomes convex. This inverse-S-shape indicates that option traders are more sensitive to the change in probability of realizing a “big loss" and “big profit"; in other words, their behavior is such that option traders are more fearful than spot traders.

\normalem

\end{spacing}
\end{document}